\documentclass[12pt]{iopart}
\usepackage{latexsym}
\usepackage{amssymb}
\usepackage{epsf}
\usepackage{xypic}
\usepackage{setspace}

\newtheorem{defn}{Definition}
\newtheorem{principle}{Principle}
\newcommand{\cA}{\mathcal{A}}
\newcommand{\cHb}{\mathcal{H}}

\newcommand{\bA}{\mathbb{A}}
\newcommand{\bC}{\mathbb{C}}

\newcommand{\bZ}{\mathbb{Z}}

\newcommand{\ass}{\mathfrak{a}}
\newcommand{\qss}{\mathfrak{q}}
\newcommand{\com}{\mathfrak{c}}
\newcommand{\lid}{\mathfrak{l}}
\newcommand{\mfi}{\mathfrak{i}}
\newcommand{\rid}{\mathfrak{r}}

\newcommand{\ph}[1]{\phantom{#1}}

\newcommand{\URep}{\textrm{URep}}
\newcommand{\piRW}{\pi \textrm{RW}}
\newcommand{\can}{\textrm{{\bf can}}}

\newcommand{\Irr}{\textrm{Irr}}
\newcommand{\Res}{\textrm{Res}}

\begin{document}

\title{Quark State Confinement as a Consequence of the Extension of the
  Bose--Fermi Recoupling to $SU(3)$ Colour}
\author{William P. Joyce}
\address{Department of Physics \& Astronomy, University of Canterbury,
  Private Bag 4800, Christchurch, New Zealand.}

\begin{abstract}
  The Bose--Fermi recoupling of particles arising from the
  $\bZ_{2}$--grading of the irreducible representations of $SU(2)$ is
  responsible for the Pauli exclusion principle. We demonstrate from
  fundamental physical assumptions how to extend this to gradings, other
  than the $\bZ_{2}$ grading, arising from other groups. This requires
  non--associative recouplings where phase factors arise due to {\it
    rebracketing} of states. In particular, we consider recouplings for
  the $\bZ_{3}$--grading of $SU(3)$ colour and demonstrate that all the
  recouplings graded by triality leading to the Pauli exclusion
  principle demand quark state confinement. Note that quark state
  confinement asserts that only ensembles of triality zero are possible,
  as distinct from spatial confinement where particles are confined to a
  small region of space by a {\it confining force} such as given by the
  dynamics of QCD.
\end{abstract}

\maketitle

\section{Introduction}                                               %

Bose--Fermi recoupling leads directly to the Pauli exclusion principle
which, for example, underlies the stability of atoms. Observational
evidence shows that particles come either as bosons or fermions.
Particle statistics arise from the phases associated with the recoupling
of states. A recoupling for the representations of $SU(2)$, where a sign
change is introduced for interchange of half integer spin and no sign
change for interchange involving an integer spin, generates
symmetric boson and anti--symmetric fermion states.

In the early days of the quark model it was realised that certain
fermionic particle resonances, based on conventional reasoning,
appeared to have symmetric states. An example, given in Kaku \cite{mk},
is the resonance $\Delta^{++}$ composed of three up quarks of total spin
$\frac{3}{2}$. The state must be symmetric in quark flavour and the spin
$\frac{1}{2}$ of each quark must be aligned. The state must also be
symmetric in quark spin. Hence the overall state is symmetric, yet the
resonance is fermionic. The solution was to introduce $SU(3)$ colour to
generate anti--symmetric quark states. Although the existence of quarks
is well established, a single free quark has never been observed. This
we call quark state confinement. We distinguish this from spatial
confinement which accounts for the localisation of quarks to a small
region of space. The latter arises from the dynamics of a theory such as
QCD. We argue that quark state confinement is a result of any $\bZ_{3}$
graded recoupling for $SU(3)$ colour admitting Pauli exclusion of
quarks. Furthermore, we determine exactly when a generalised Bose-Fermi
grading leads to state confinement.

We consider physical systems conforming to the following assumptions:
\begin{enumerate}
\item{The system possesses an exact symmetry given by some
    semi--simple group $G$}
\item{Single particle state spaces are finite unitary irreducible
    representations of the group $G$.}
\item{Composite (particle) state spaces are given by coupling together
    consistuent single particle state spaces using tensor product.}
\item{Recoupling of composite state spaces is a natural isomorphism.}
\end{enumerate}
The first three assumptions are well established quantum axioms. The
fourth perhaps needs some expanation. A recoupling is an invertible
intertwiner ($G$--equivariant, unitary and linear) satisfying a
naturality condition. Naturality is an important idea coming from
category theory \cite{sm}. For example, given three particles with state
spaces $\cHb_{1}$, $\cHb_{2}$ and $\cHb_{3}$ in the state $\psi_{1}$,
$\psi_{2}$ and $\psi_{3}$ respectively, a recoupling between the
physically equivalent state spaces $(\cHb_{1}\otimes \cHb_{2})\otimes
\cHb_{3}$ and $\cHb_{2}\otimes (\cHb_{1}\otimes \cHb_{3})$ is an
invertible interwtiner $T_{\cHb_{1},\cHb_{2},\cHb_{3}}:(\cHb_{1}\otimes
\cHb_{2})\otimes \cHb_{3}\rightarrow \cHb_{2}\otimes (\cHb_{1}\otimes
\cHb_{3})$ recoupling the states $(\psi_{1}\otimes \psi_{2})\otimes
\psi_{3}$ to something like $\psi_{2}\otimes (\psi_{1}\otimes
\psi_{3})$. The natural condition satisfied is, given any individual
observation or preparation of the individual states by linear operators
$A_{i}:\cHb_{i}\rightarrow\cHb^{\prime }_{i}$ changing the state
$\psi_{i}$ to $\psi^{\prime }_{i}$ then the following diagram
commutes\\ \\
\hspace*{2cm}
\xymatrix@=60pt{(\cHb_{1}\otimes \cHb_{2})\otimes
  \cHb_{3}\ar[r]^{T_{\cHb_{1},\cHb_{2},\cHb_{3}}} \ar[d]_{(A_{1}\otimes
    A_{2})\otimes A_{3}} & \cHb_{2}\otimes
  (\cHb_{1}\otimes \cHb_{3}) \ar[d]^{A_{2}\otimes (A_{1}\otimes A_{3})}\\
  (\cHb^{\prime }_{1}\otimes \cHb^{\prime }_{2})\otimes \cHb_{3}
   \ar[r]^{T_{\cHb^{\prime }_{1},\cHb^{\prime }_{2},\cHb^{\prime
         }_{3}}}
   & \cHb^{\prime }_{2}\otimes (\cHb^{\prime }_{1}\otimes \cHb^{\prime
    }_{3})}\\ \\
Normally for $SU(2)$ with Bose--Fermi recoupling the horizontal arrows
introduce no phase change, but as we shall see this is not the case for
$SU(3)$ colour Bose--Fermi recoupling.

There is a long history of investigation into associative recoupling,
beginning with the early work of Green \cite{hg}. Green generalised
quantisation of associative algebras of annihilation and creation
operators. Such generalisations led to parastatistics \cite{abhg, hg3,
  hgpj, wm}, modular statistics \cite{hg2} and graded Lie algebras
\cite{lcynss, ms}. These approaches all work with algebras having an
associative universal embedding algebra and have been used to describe
some features of the quark model. However, this approach has not been
able to explain confinement, instead arguing that its origin is dynamical.

In this paper we do not restrict ourselves to associative recoupling.
Instead we seek the most general recoupling consistent with the physical
requirements of a quantum system exhibiting symmetry. Furthermore, we
make no assumptions about the existence of a generalised colour algebra
nor attempt to explain the quark model. We simply determine the
ramifications of a Bose--Fermi recoupling for $SU(3)$ colour. The
non--associativity is required to accommodate Bose--Fermi recouplings
over a $\bZ_{3}$--gradation. There is no physical reason why
non--associative recouplings are not admissable. In fact the statistical
consequence is quark state confinement without taking into consideration
dynamics. These results were announced in Joyce \cite{wj4}.

A symmetric monoidal structure of the category of unitary
representations provides a framework for describing recoupling, and the
Racah--Wigner calculus. We refer the reader elsewhere for an
introduction to category theory, group representation theory and the
Racah--Wigner calculus. The book by Mac Lane \cite{sm} is the standard
reference on category theory. An introduction to braided monoidal
categories in the context of quantum groups are Kassel \cite{ck} and
Majid \cite{smj}. The group representation notation used in this paper
is based on Br\"ocker and tom Dieck \cite{tbtd}. A gentle introduction
to a category theoretic formulation of the Racah--Wigner calculus is
given in Joyce {\it et.  al.} \cite{wjpbhr} and for coupling theory
Joyce \cite{wj2}. Although category theory is the best language to
describe recoupling, we trust that much of the paper is accessable
through examples, and the useage of non--categorical language whenever
it is feasible to do so.

We demonstrate in this paper that a Bose--Fermi colour recoupling is
neither a symmetric monoidal nor a braided monoidal structure. Colour
recoupling requires a symmetric premonoidal structure as defined in
Joyce \cite{wj, wj3}. A symmetric premonoidal structure introduces a
natural automorphism to account for the non--commutativity of the
pentagon diagram. Hence recouplings based on symmetric premonoidal
structures is necessary and leads to a deformation/generalisation of the
usual Racah--Wigner calculus. This calculus together with appropriate
diagram notation is developed in a series of papers \cite{wjI,
  wjII, wjIII, wjIV, wjV}.

\section{Recoupling and Statistics}                                  %

The collection of unitary representations for a group $G$ is a symmetric
monoidal category $\URep_{G}$. Loosely it is equiped with a tensor
product and recoupling structure. Let $\Irr_{G}$ denote a
collection of isotypical irreducible representations (or irreps).
Suppose that $G$ is semi--simple so that every representation is decomposible
as a direct sum of elements from $\Irr_{G}$. A one particle ket state of
the system is the mapping
\begin{eqnarray}
|\psi \rangle :\bC \rightarrow (\lambda )
\end{eqnarray}
given by $z\mapsto z\psi $ where $z\in \bC $, $\lambda \in \Irr_{G}$,
the round brackets is the restriction functor $\Res^{G}$ taking $\lambda
\mapsto (\lambda )=\Res^{G}\lambda =\bC^{|\lambda |}$ and $\psi \in
(\lambda )$. One should think of $(\lambda )$ as the state space of the
particle described by the irrep $\lambda $. For example, the spin half
irrep's state space is two dimensional and spanned by basis vectors
corresponding to spin up and spin down along some axis. Multi--particle
states are formed by ``tensoring'' single particle states together. The
irreps, under tensor product, generate the (projected) Racah--Wigner
category $\piRW_{G}$.  This category inherits the symmetric monoidal
structure of $\URep_{G}$.

Multi--particle states are built out of single particle states, the
state space being given by the tensor product of the single particle
states. Given $n$ particles contained in $n$ irreps, the state space
representing this multi--particle system is dependent on the order and
bracketing of irreps. A particular choice is called an ensemble. We
abuse notation and call each irrep a particle. The natural isomorphisms
of the symmetric monoidal structure reorder and rebracket ensembles. The
order in which the $n$ irreps are coupled is represented by a rooted planar
binary tree with labeled leaves. This is called a bracketing tree, see
Joyce \cite{wj2}. Operations between bracketing trees are called
recouplings.

The recoupling between ensembles is given by the symmetric monoidal
structure of $\piRW_{G}$. That is, by associativity($\ass $),
commutativity($\com $) and left and right identity($\lid $ and $\rid $)
natural isomorphisms, where we denote the identity irrep by $e$. These
determine respectively natural isomorphisms
\begin{eqnarray}
\ass_{a,b,c} :(a\otimes b)\otimes c & \rightarrow & a\otimes (b\otimes c)\\
\ph{spaces}\com_{a,b}:a\otimes b &\rightarrow & b\otimes a\\
\ph{spacesp}\lid_{a}:e\otimes a & \rightarrow & a\\
\ph{spacesp}\rid_{a}:a\otimes e & \rightarrow & a
\end{eqnarray}
representing rebracketing, adjacent transposition and the removal of
the vacuum from the left or right. Given any two couplings of a set of
irreps there are a number of differing sequences of the above
elementary recouplings transforming one into the other. If these two
sequences compose to always give the same natural isomorphism we say
that the structure is coherent. The Mac Lane coherence theorem
\cite{sm, sm2} asserts that a necessary and sufficient condition for
coherence is that the pentagon, hexagon and triangle diagrams commute
and that commutativity is symmetric. The symmetry of commutativity
asserts $\com_{b,a}=\com^{-1}_{a,b}$. The pentagon diagram is\\ \\
\hspace*{2cm}
\xymatrix@=40pt{
((a\otimes b)\otimes c)\otimes d \ar[r]^{\ass_{a\otimes b,c,d}}
\ar[d]_{\ass_{a,b,c}\otimes 1_{d}} &
(a\otimes b)\otimes (c\otimes d) \ar[r]^{\ass_{a,b,c\otimes d}} &
a\otimes (b\otimes (c\otimes d)) \\
(a\otimes (b\otimes c))\otimes d \ar[rr]_{\ass_{a,b\otimes c,d}} & &
a\otimes ((b\otimes c)\otimes d) \ar[u]_{1_{a}\otimes \ass_{b,c,d}}
}\\ \\
The hexagon diagram is\\ \\
\hspace*{2cm}
\xymatrix@=40pt{
(a\otimes b)\otimes c \ar[r]^{\ass_{a,b,c}} \ar[d]_{\com_{a,b}\otimes 1_{c}} &
a\otimes (b\otimes c) \ar[r]^{\com_{a,b\otimes c}} &
(b\otimes c)\otimes a \ar[d]^{\ass_{b,c,a}} \\
(b\otimes a)\otimes c \ar[r]_{\ass_{b,a,c}} &
b\otimes (a\otimes c) \ar[r]_{1_{b}\otimes \com_{a,c}} &
b\otimes (c\otimes a)
}\\ \\
Lastly the triangle diagram is\\ \\
\hspace*{2cm}
\xymatrix{
(a\otimes e)\otimes b \ar[rr]^{\ass_{a,e,b}}
\ar[dr]_{\rid_{a}\otimes 1_{b}} & &
a\otimes (e\otimes b) \ar[dl]^{1_{a}\otimes \lid_{b}}\\
& a\otimes b &
}\\ \\

We require the states of any composite system to be compatible with the
recoupling structure. That is, given a state $|\psi \rangle :\bC
\rightarrow (a)$ and an automorphic recoupling $\mfi :a\rightarrow a$
then the following diagram is commutative.\\ \\
\hspace*{2cm}
\xymatrix@=30pt{(a)\ar[r]^{(\mfi )} & (a) \\
  \bC \ar[u]^{|\psi \rangle } \ar[ur]_{\tau_{\pi }|\psi \rangle } &
  }\\ \\
where $\pi $ is the permutation of particles given by $\mfi $ and
$\tau_{\pi }(a_{1}\otimes \cdots \otimes a_{n})=a_{\pi 1}\otimes \cdots
\otimes a_{\pi n}$. The map $i:a\rightarrow a$ represents the recoupling
of identical particles by permuting amongst themselves their order in
the ensemble $a$.  Alternatively, given any map $|\psi \rangle :\bC
\rightarrow (a)$ then a state of the system is given by
\begin{eqnarray}
\sum_{\mfi }(\mfi )|\psi \rangle
\end{eqnarray}
where we sum over all recouplings $\mfi :a\rightarrow a$. If the
particle labels of $a$ are all distinct then the only recoupling is
the identity.

We define an equivalence on the set of ensembles given by $a\sim b$ if
and only if there is an ensemble $c$ such that $a$ and $b$ are contained
in the direct sum decomposition of $c$ (written $a,b\subset c$). In
other words the ensemble $c$ may interact in some way to become either
$a$ or $b$ (ignoring dynamical and kinematic considerations). The
set of equivalence classes $[a]=\{ b:a\sim b\} $ forms an Abelian group
$\bA $ with addition $[a]+[b]=[a\otimes b]$ and identity $0=[e]$. The
inverse of $[a]$ is given by $-[a]=[a^{*}]$ since $e\subset a\otimes
a^{*}$. To give some examples, if $G=SU(n)$ then $\bA =\bZ_{n}$. If
$G=C_{n}$ and $n\geq 1$ then $\bA =\bZ_{2n}$. If $G=SO(3)$ then $\bA
=\bZ_{1}$.  If $G=D_{n}$ where $n\geq 2$, or $G$ is the tetrahedral,
octahedral or icoshedral group then $\bA =\bZ_{2}$.

The natural square property of the recouplings mapped under the
restriction functor are required to be natural at the state level.
This allows us to conclude that the recouplings are of the form
\begin{eqnarray}
\ass_{a,\b,c}\Big( (a_{i}\otimes b_{j})\otimes c_{k}\Big) & = &
\alpha_{m,n,p}a_{i}\otimes (b_{j}\otimes c_{k}) \\
\ph{spacesp}\com_{a,b}\Big( a_{i}\otimes b_{j}\Big) & = & \gamma_{m,n}b_{j} 
\otimes a_{i} \\
\ph{spacespac}\lid_{a} \Big( e\otimes a_{i}\Big) & = & \lambda_{m}a_{i} \\
\ph{spacespac}\rid_{a} \Big( a_{i}\otimes e\Big) & = & \rho_{m}a_{i}
\end{eqnarray}
where $m=[a]$, $n=[b]$, $p=[c]$, $\{ a_{i}\}_{i}$ is a basis for $a$,
$\{ b_{j}\}_{j}$ is a basis for $b$ and $\{ c_{k}\}_{k}$ is a basis for
$c$. See the appendix for details. The pentagon, hexagon, symmetry and
triangle conditions place the following constraints on the phases.
\begin{eqnarray}
\ph{sp}\alpha_{m+n,p,q}\alpha_{m,n,p+q} & = &
\alpha_{m,n,p}\alpha_{m,n+p,q}\alpha_{n,p,q} \label{pentcond}\\
\alpha_{m,n,p}\gamma_{m,n+p}\alpha_{n,p,m} & = &
\gamma_{m,n}\alpha_{n,m,p}\gamma_{m,p} \label{hexcond}\\
\ph{spacespa}\gamma_{m,n}\gamma_{n,m} & = & 1 \label{symcond}\\
\ph{spacespa}\alpha_{m,0,n}\lambda_{n} & = & \rho_{m} \label{tricond}
\end{eqnarray}
Any choice of phase factors satisfying these conditions defines a
recoupling. We give some examples:
\begin{enumerate}
\item{We have the (pure) Bose recoupling where all phases are unity. If
    $\bA =\bZ_{1}$ the only recoupling is Bose recoupling.}
\item{If $\bA=\bZ_{2}$ the Bose--Fermi recoupling is given by
    $\gamma_{1,1}=-1$. All other phases must be unity. Compatibility of
    states with this recoupling leads to symmetric states for bosons
    (even grade) and anti--symmetric states for fermions (odd grade).
    From this follows the Pauli exclusion principle.}
\item{If $\bA =\bZ_{n}$ then by the hexagon condition (\ref{hexcond})
    associative recouplings satisfy
    $\gamma_{m+p,q}=\gamma_{m,q}\gamma_{p,q}$. The general solution is
    easily found by induction to be $\gamma_{p,q}=(\gamma_{1,1})^{pq}$.
    The symmetry condition (\ref{symcond}) gives $\gamma_{1,1}=\pm 1$.
    Hence there are only two associative recouplings. Given an $\bA
    $--graded associative algebra $\cA =\oplus_{m\in \bA }A_{m}$ one may
    construct the bracket
\begin{eqnarray}
[a,b] & = & ab-\gamma_{n,m}ba
\end{eqnarray}
satisfying $[b,a]=-\gamma_{m,n}[a,b]$ where $a\in \cA_{m}$ and $b\in
\cA_{n}$. This bracket satisfies the Jacobi identity
$[a,[b,c]]=[[a,b],c]+\gamma_{n,m}[b,[a,c]]$ where $c\in \cA_{p}$. The
algebra $A$ with this bracket is a Lie algebra for $\gamma_{1,1}=1$ and
a graded Lie algebra for $\gamma_{1,1}=-1$.}
\item{If we have recouplings $\alpha_{m,n,p}$, $\gamma_{m,n}$,
    $\lambda_{m}$ and $\rho_{n}$ for $m,n,p\in \bA $, and
    $\alpha^{\prime }_{m^{\prime },n^{\prime },p^{\prime }}$,
    $\gamma^{\prime }_{m^{\prime },n^{\prime }}$, $\lambda^{\prime
      }_{m^{\prime }}$ and $\rho^{\prime }_{n^{\prime }}$ for $m^{\prime
      },n^{\prime },p^{\prime }\in \bA^{\prime }$ then the point--wise
    product $\alpha_{m,n,p}\alpha^{\prime }_{m^{\prime },n^{\prime
        },p^{\prime }}$, $\gamma_{m,n}\gamma^{\prime }_{m^{\prime
        },n^{\prime }}$, $\lambda_{m}\lambda^{\prime }_{m^{\prime }}$
    and $\rho_{m}\rho^{\prime }_{n^{\prime }}$ is a recoupling for $\bA
    \times \bA^{\prime }$.}
\end{enumerate}

In QCD one would like to introduce $SU(3)$ colour and require that it
carries a Bose--Fermi recoupling. However, $\bA =\bZ_{3}$ obstructs the
recoupling from being a symmetric or braided monoidal structure.  Let
$1$ be the class containing the $SU(3)$ representation $[3]$ and $2$ its
dual $\overline{[3]}$. We require $\gamma_{1,1}=\gamma_{2,2}=-1$. A
symmetric monoidal recoupling requires $\gamma_{2,2}=1$ as we now show.
The hexagon condition (\ref{hexcond}) with $m=n=p=1$ gives
$\alpha_{1,1,1}\gamma_{1,2}\alpha_{1,1,1}=\gamma_{1,1}^{2}\alpha_{1,1,1}$.
Thus the symmetry condition (\ref{symcond}) implies
$\gamma_{2,1}=\alpha_{1,1,1}$. This together with the hexagon condition
(\ref{hexcond}) with $m=2$ and $n=p=1$ gives
\begin{eqnarray}
\gamma_{2,2} & = &
\frac{\alpha_{1,1,1}^{2}\alpha_{1,2,1}}{\alpha_{2,1,1}\alpha_{1,1,2}}
\label{apentcond}
\end{eqnarray}
But the pentagon condition (\ref{pentcond}) with $m=n=p=q=1$ implies
that $\gamma_{2,2}=1$. Hence the colour recoupling cannot be a symmetric
monoidal recoupling. Even though such a recoupling may be
non--associative, it is to restrictive. Two possibilities exist: a
braided monoidal recoupling (see Joyal and Street \cite{ajrs}) or a
symmetric premonoidal recoupling (see Joyce \cite{wj, wj3}). However,
the braided monoidal recoupling cannot describe the colour recoupling
because the second hexagon equation with $m=n=p=1$ and the requirement
$\gamma_{1,1}^{2}=1$ shows $\gamma_{1,2}=\alpha_{1,1,1}$. But from the
first hexagon we have that $\gamma_{1,2}^{-1}=\alpha_{1,1,1}$. Thus
$\gamma_{1,2}^{2}=\alpha_{1,1,1}^{2}=1$. Similarly one deduces that
$\gamma_{2,1}^{2}=\alpha_{2,2,2}^{2}=1$. Importantly, the pentagon
condition (\ref{pentcond}) above shows that $\gamma_{2,2}=1$. There is,
however, an important reason why a braid must be symmetric. If we apply
commutative recoupling twice to a state $|\psi \rangle:\bC \rightarrow
(a\otimes b)$ we see that $|\psi \rangle =\gamma_{b,a}\gamma_{a,b}|\psi
\rangle $ which only admits non--trivial solutions when the symmetry
condition (\ref{symcond}) holds. Only a symmetric premonoidal recoupling
is capable of describing a colour recoupling as we demonstrate in the
next section.

\section{Symmetric Premonoidal Recoupling}                           %

We begin by carefully revisiting the notion of coupling. A coupling tree
is a rooted planar binary tree with a linear ordering of its vertices
such that every shortest path from the root to a leaf is an increasing
sequence and a linear ordering of its leaves. An example is given in
figure 2.
\begin{figure}[h]
\hspace*{2cm}
\epsfxsize=200pt
\epsfbox{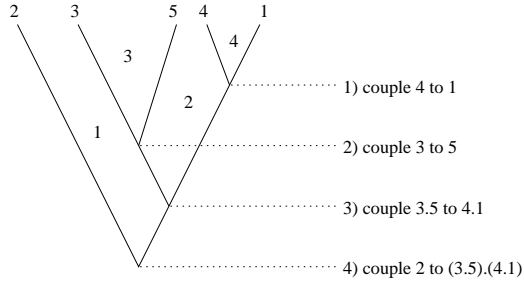}
\caption{An example of a coupling tree.}
\end{figure}
One should note that the level of the vertices in the tree determines
the coupling hierachy. In this example the coupling sequence is $1324$.
An ensemble tree is given by evaluation by irrep labels. Given a tuple
of labels, we label the leaf in the $i$th poistion of the linear
ordering by the labeled $l_{i}$. The recouplings are represented by
unique arrows between coupling trees characterised by a pair of
permutations. Note that many coupling trees evaluate to the same
ensemble tree. The canonical functor $\can $ maps ensemble trees to
ensembles and recouplings to natural isomorphisms in the obvious way. An
example is given in figure 3.
\begin{figure}[h]
\hspace*{2cm}
\epsfxsize=240pt
\epsfbox{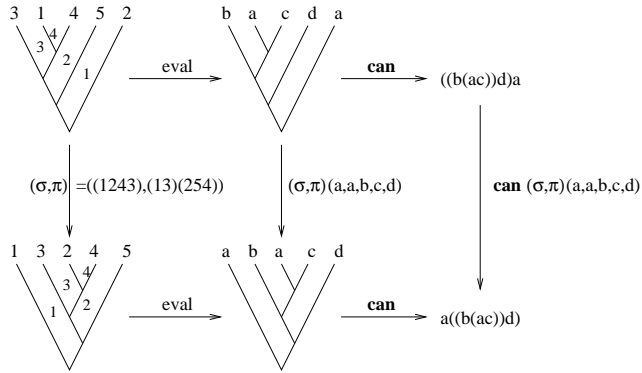}
\caption{An example of the recoupling $\sigma $ between two coupling
  trees, their evaluation by $(a,a,b,c,d)$ to ensemble trees and
  subsequent mapping under $\can $ to ensembles.}
\end{figure}
The ensemble tree represents physically distinct coupling scenarios
that take into account particle indistinguishability. The coupling
trees serve to distinguish recouplings and the ensembles are the state
spaces. The permutation $\sigma $ permutes the coupling sequence, the
permutation $\pi $ permutes the order of the particles. For a
comprehensive exposition see Joyce \cite{wj3, wjI}.

We introduce a deformativity natural automorphism $\qss $ to represent
the non--commutativity of the pentagon diagram. This is depicted in figure 1.
\begin{figure}[h]
\hspace*{2cm}
\epsfxsize=240pt
\epsfbox{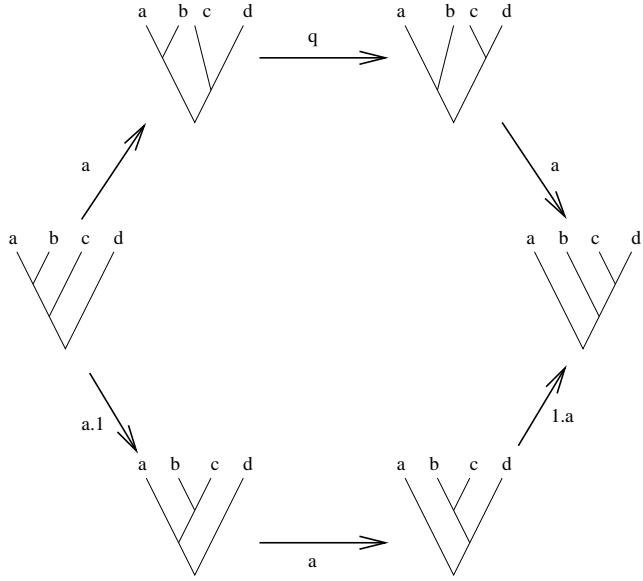}
\caption{The $\qss $--pentagon diagram of a premonoidal structure, where
  $\qss $ represents the degree to which the pentagon diagram does not
  commute.}
\end{figure}
Thus for example, in the ensemble $(a\otimes b)\otimes (c\otimes d)$ we
distinguish between coupling $a$ to $b$ before, as opposed to after,
coupling $c$ to $d$. The functor $\can $ is coherent
if the hexagon diagram and triangle diagrams commute, and the
following three diagrams commute.\\ \\
\xymatrix{
(e\otimes a)\otimes b \ar[rr]^{\ass_{e,a,b}}
\ar[dr]_{\rid_{a}\otimes 1_{b}} & &
e\otimes (a\otimes b) \ar[dl]^{\rid_{a\otimes b}}\\
& a\otimes b & }\hspace*{1cm}
\xymatrix{
(a\otimes b)\otimes e \ar[rr]^{\ass_{a,b,e}}
\ar[dr]_{\lid_{a\otimes b}} & &
a\otimes (b\otimes e) \ar[dl]^{1_{a}\otimes \lid_{b}}\\
& a\otimes b & }\\ \\
\hspace*{4cm}
\xymatrix@=40pt{
(a\otimes b)\otimes (c\otimes d) \ar[r]^{\qss_{a,b,c,d}}
\ar[d]_{\com_{a\otimes b,c\otimes d}} & (a\otimes b)\otimes (c\otimes
d) \ar[d]_{\com_{a\otimes b,c\otimes d}} \\
(c\otimes d)\otimes (a\otimes b) & (c\otimes d)\otimes (a\otimes b)
\ar[l]_{\qss_{c,d,a,b}}}\\ \\

The deformativity recoupling is given by (see the appendix)
\begin{eqnarray}
\qss \Big( (a_{i}\otimes b_{j})\otimes (c_{k}\otimes d_{l})\Big) & = &
\xi_{a,b,c,d}(a_{i}\otimes b_{j})\otimes (c_{k}\otimes d_{l})
\end{eqnarray}
where $\xi_{a,b,c,d}$ is a phase factor and a class function of the
$\bA $--gradation. The constraints on the recoupling phases are
\begin{eqnarray}
\alpha_{m+n,p,q}\xi_{m,n,p,q}\alpha_{m,n,p+q} & = &
\alpha_{m,n,p}\alpha_{m,n+p,q}\alpha_{n,p,q} \label{qpent}\\
\ph{spac}\alpha_{m,n,p}\gamma_{m,n+p}\alpha_{n,p,m} & = &
\gamma_{m,n}\alpha_{n,m,p}\gamma_{m,p} \label{hex}\\
\ph{spacespa}\xi_{m,n,p,q}\xi_{p,q,m,n} & = & 1 \label{sqr}\\
\ph{spacespacesp}\gamma_{m,n}\gamma_{n,m} & = & 1 \label{sym}\\
\ph{spacespace}\alpha_{0,m,n}\lambda_{m+n} & = & \lambda_{m} \label{ltri}\\
\ph{spacespacesp}\alpha_{m,0,n}\lambda_{n} & = & \rho_{m} \label{tri}\\
\ph{spacespacesp}\alpha_{m,n,0}\rho_{n} & = & \rho_{m+n} \label{rtri}
\end{eqnarray}
for all $m,n,p,q\in \bA $. Note that (\ref{qpent}) provides a formula
for $\xi_{m,n,p,q}$. Let $S^{1}=\{ z\in \bC :|z|=1\} \subset \bC $ be 
the set of phase factors. We now give a formal definition of a
recoupling for an Abelian group $\bA $.
\begin{defn}
A recoupling for an Abelian group $\bA $ consists of the four
maps $\alpha :\bA^{3}\rightarrow S^{1}$, $\gamma :\bA^{2}\rightarrow
S^{1}$ and $\lambda ,\rho :\bA \rightarrow S^{1}$ satisfying conditions
(\ref{qpent}) through (\ref{rtri}).
\end{defn}
A recoupling is called a Bose--Fermi recoupling whenever
$\gamma_{m,m}=-1$ for all $m\in \bA \setminus \{ 0\} $.
We can define a Bose--Fermi recoupling for any $\bA $--gradation as
follows. We take $\lambda_{m}=\rho_{n}=1$ and
\begin{eqnarray}
\gamma_{m,n} & = & \left\{ \begin{array}{cl} 1 & :m=0 \textrm{ or } n=0 \\ 
    -1 & :\textrm{otherwise}\end{array} \right. \\
\alpha_{m,n,p} & = & \left\{ \begin{array}{cl} 1 & :m=0, n=0, p=0
    \textrm{ or } m+n=0 \\ 
    -1 & :\textrm{otherwise}\end{array} \right.
\end{eqnarray}
The $m+n=0$ in the definition of $\alpha_{m,n,p}$ may equally well be
replaced by $n+p=0$. These determine the deformativity phases to
be
\begin{eqnarray}
\xi_{m,n,p,q} & = & \left\{ \begin{array}{cl} 1 & :m=0, n=0, p=0, q=0, m+n=0
    \textrm{ or } p+q=0 \\ 
    -1 & :\textrm{otherwise}\end{array} \right.
\end{eqnarray}
We immediately see that the recoupling is monoidal for $\bA =\bZ_{2}$,
but premonoidal for $\bA =\bZ_{n}$ where $n\geq 3$. To verify the phase
conditions we only need to demonstrate the hexagon condition (\ref{hex})
holds and that the definition of $\xi_{m,n,p,q}$ is correct, the other
conditions are immediate. If $m=0$, $n=0$ or $p=0$ it is easily shown.
Suppose they are all non--zero then $\gamma_{m,n}\gamma_{m,p}=1$.  If
$n+p=0$ then the hexagon condition reduces to
$\alpha_{m,n,-n}\alpha_{n,-n,m}=\alpha_{n,m,-n}$ which holds.  Now
suppose also that $n+p\neq 0$ then $\gamma_{m,n+p}=-1$ and the hexagon
condition is $\alpha_{m,n,p}\alpha_{n,p,m}=-\alpha_{n,m,p}$ which holds.
A similar argument shows the definition of $\xi_{m,n,p,q}$ is correct.

For this $\bZ_{3}$--graded Bose--Fermi recoupling all phases are unity
except the following which are $-1$.
\begin{eqnarray}
\gamma_{1,1}\ph{space} & \alpha_{1,1,1}\ph{space} & \xi_{1,1,1,1}\nonumber \\
\gamma_{1,2}\ph{space} & \alpha_{1,1,2}\ph{space} & \xi_{1,1,2,2} \\
\gamma_{2,1}\ph{space} & \alpha_{2,2,1}\ph{space} & \xi_{2,2,1,1}\nonumber \\
\gamma_{2,2}\ph{space} & \alpha_{2,2,2}\ph{space} & \xi_{2,2,2,2} \nonumber
\end{eqnarray}

\section{Exclusion and Confinement Principles}                       %

Given an ensemble of particles, sometimes there are a number of coupling
schemes associated with it. This occurs when there are identical
particles, or when the coupling process is non--monoidal. These
situations lead respectively to exclusion and confinement
principles.

Indistinguishability requirements place statistical constraints on what states
of a given system are possible. Given an ensemble tree $w$ the state space
of the system is $\cHb =(\can w)$. Thus a map $|\psi \rangle
\rightarrow \cHb $ is a state of the system if it is compatible with
the two following conditions.
\begin{enumerate}
\item{{\bf Indistinguishability of particles:} Given ensemble trees $w$
    and $w^{\prime }$ with the same state space $\cHb $ then $|\psi
    \rangle :\bC \rightarrow \cHb $ is a state of the system if for every
    recoupling $(\sigma ,\pi ):w\rightarrow w^{\prime }$ the diagram below
    commutes.\\ \\
\hspace*{2cm}
\xymatrix@=40pt{\cHb \ar[r]^{(\can (\sigma ,\pi ))} & \cHb \\
\bC \ar[u]^{|\psi \rangle } \ar[ur]_{\tau_{\pi }|\psi \rangle } & }\\ \\
where $\tau_{\pi }(a_{1}\otimes \cdots \otimes a_{n})=a_{\pi 1}\otimes
\cdots \otimes a_{\pi n}$.}
\item{{\bf Composition of particles:} Given two states $|\psi \rangle :\bC
    \rightarrow \cHb $ and $|\psi^{\prime }\rangle :\bC \rightarrow \cHb 
    $ the composite $|\Psi \rangle :\bC \rightarrow \cHb $ given by the
    commuting of the diagram below is a state (and so satisfies (i)).\\ \\
\hspace*{2cm}
\xymatrix@=40pt{\bC^{2}\ar[r]^{|\psi \rangle \otimes |\psi^{\prime
      }\rangle \ph{spa}} & \cHb \otimes \cHb^{\prime }\\
\bC \ar[u]^{\Delta } \ar[ur]_{|\Psi \rangle } & }\\ \\
    where $\Delta z=(z,z)$ for all $z\in \bC $ is the diagonal map.}
\end{enumerate}
Note that if the recoupling is symmetric monoidal then
property (ii) follows from (i).

The next result deduces the generalisation of the Pauli exclusion principle.
\begin{principle}(Exclusion) Given an ensemble of identical particles $a$,
  the Bose--Fermi recoupling asserts that the state is
  symmetric if $a\in 0$ and anti--symmetric otherwise.
\end{principle}
This justifies the name of the recoupling and is the Pauli exclusion
principle for $G=SU(2)$.\\
{\it Proof:} Given any coupling tree $w$ we wish to determine a sequence
of associativity and one commutativity recouplings the interchange the
$i$th and $i+1$th leaf. To do this determine a sequence of associativity
recouplings that ensures the $i$th and $i+1$th leafs are coupled
together first in the coupling tree. Next apply the commutativity
recoupling swaping them, and finally reverse the sequence of
associaitivty recouplings to give a coupling tree $w^{\prime }$ that
only differs from $w$ by the interchange of the $i$th and $i+1$th leaves.
This is depicted in figure 2.
\begin{figure}[h]
\hspace*{2cm}
\epsfxsize=240pt
\epsfbox{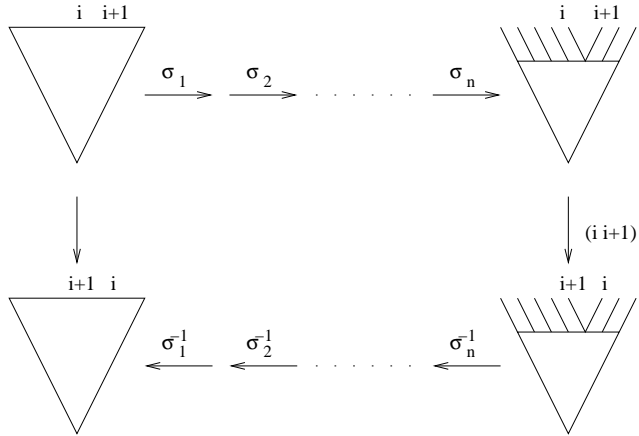}
\caption{Transposition of two adjacent particles.}
\end{figure}
Next evaluate these trees for a fixed label $a$. They give rise to the
same ensemble tree, and under $\can $ the same ensemble. The
recoupling phase is given by $\gamma_{a,a}$ since all the
associativity recoupling phases must cancel by construction. Thus any
state under adjacent interchange introduces a phase factor
$\gamma_{a,a}$. Hence by indistiguishability a state of the system is
symmetric if $\gamma_{a,a}=1$ and anti--symmetric for
$\gamma_{a,a}=-1$. $\square $

We now deduce the principle of state confinement.
\begin{principle}(Confinement) Given a Bose--Fermi recoupling then there 
  is a nilpotent $n$ of $\bA $ (that is $2n=0$) such that the non--zero
  states correspond to ensembles of grade zero and $n$. 
\end{principle}
If $\bA $ has no non--zero nilpotent grades the non--zero states are
confined to grade zero ensembles. This is the situation for $SU(3)$
colour giving quark state confinement.\\
{\it Proof:} We begin by proving that
\begin{eqnarray}
\xi_{m,n,m,n} & = & \gamma_{m+n,m+n}\gamma_{m,m}\gamma_{n,n}
\end{eqnarray}
The hexagon condition (\ref{hex}) gives
$\alpha_{m+n,m,n}\alpha_{m,n,m+n}=\alpha_{m,m+n,n}\gamma_{m+n,m+n}
\gamma_{m+n,m}\gamma_{m+n,n}$.  Substituting this into the formula
(\ref{qpent}) for $\xi_{m,n,m,n}$ gives
$\xi_{m,n,m,n}=\alpha_{m,n,m}\alpha_{n,m,n}\gamma_{m+m,m+n}$
$\gamma_{m,m+n}\gamma_{n,m+n}$.  Again the hexagon condition (\ref{hex})
gives $\alpha_{m,n,m}=\gamma_{m,n}\gamma_{m,m}\gamma_{m+n,m}$, and a
similar formula with $m$ and $n$ interchanged. Substituting these into
the previous expression gives the desired formula.  If $a$ corresponds
to an ensemble for which its grade $[a]=m$ does not generate $\bZ_{1}$
or $\bZ_{2}$ then $\xi_{m,m,m,m}=\gamma_{2m,2m}=-1$. Now the composition
of state property applied to a state $|\psi \rangle :\bC \rightarrow
(a)$ gives the $4$--fold composite state $|\Psi \rangle :\bC \rightarrow
((a\otimes b)\otimes (a\otimes b))$ satisfying $|\psi \rangle
=\xi_{m,n,m,n}|\psi \rangle $. This can only occur if $|\psi \rangle
=0$. The ensembles admitting non--trivial states generate an Abelian
subgroup $\bA_{0}$ of grades $m$, $n$ satisfying $m+n=0$ because if
$m+n\neq 0$ either $m$ or $n$ would admit only trivial states. Hence
$\bA_{0}$ is $\bZ_{1}$ or $\bZ_{2}$ giving the desired nilpotent. Either
way the deformativity phase is always zero. $\square $

For $SU(2)$, which is $\bZ_{2}$--graded, one arrives at the conclusion
that the only non--unity phase possible is $\gamma_{1,1}$. Moreover,
the recouplings are symmetric monoidal and there is only one choice of
Bose--Fermi recoupling ($\gamma_{1,1}=-1$). Thus Pauli exclusion
follows and there is no state confinement requirement. On the other
hand for $SU(3)$, which is $\bZ_{3}$--graded, there are a number of
Bose--Fermi recouplings. Importantly, they are all symmetric
premonoidal (never monoidal), satisfy Pauli exclusion and because of
state confinement only triality zero states are possible.

The only remaining $\bZ_{n}$--grade admitting the state confinement
observed in nature is $\bZ_{6}$. This could be aligned with $SU(6)$
flavour. However, since each quark flavour has a different mass there is 
no reason to believe that a flavour indistinguishability principle
exists. Moreover, $SU(2)$ spin and $SU(3)$ colour are sufficient to
describe the statistical behaviour observed in nature.

In standard QFT the associtivity is strict and brackets are
ignored. In other words all $\alpha_{m,n,p}$ are unity. In the case of
QCD some modification of the recoupling structure is required if
confinement is to become an intrinsic property. The only irreducible
physical ensembles are the vacuum, mesons, hadrons and free
gluons. Gluons are free to enter and exit mesons and hadrons providing
the mechanism of the strong interaction. It is important to realise
that one cannot have the Pauli exclusion principle for $SU(3)$ colour
without the confinement of quarks to mesonic and hadronic ensembles. A
formulation of many--body quantum theory taking this into account is
given in Joyce \cite{wjV}. This approach does not rely on annihilation
and creation operators. It is an open question as to what form
non--associative algebras of annihilation and creation operators might
take to accommodate non--associative recoupling.

\section{Conclusion}                                                 %

Starting from fundamental principles we derived the recoupling
structure of ensemble quantum systems with exact symmetry. This was
found to lead to a recoupling algebra of phases. The symmetry of the
situation leads to a gradation for the ensembles of which the
recoupling is a class function. There is some freedom in the
choice of phases, each leading to different statistical behaviour.

Physical requirements demand the usual Bose--Fermi recoupling over
$SU(2)$ spin and $SU(3)$ colour. In order to accommodate this for
$SU(3)$ colour we deduced the need for non--associative recoupling. More
generally we constructed a consistent Bose--Fermi recoupling for any
gradation. The recoupling algebra placed constraints on what states of
the system are allowable. For Bose--Fermi recoupling we demonstrated a
(generalised) Pauli exclusion principle holds. Additionally we proved
that a state confinement principle was unavoidable. The triality grading
of $SU(3)$ colour ensembles ensured that quark state confinement was
mandatory. No {\it confining force} was necessary to explain quark state
confinement. However, spatial confinement of quarks to within baryons is
explained by the dynamics of a theory such as QCD.

\section*{Appendix}                                                %

The natural square property of the recouplings mapped under the
restriction functor are required to be natural at the state level.
Consider commutativity then this natural
condition is as follows: Given $a\cong c$ and $b\cong d$ then\\ \\
\hspace*{2cm}
\xymatrix{(a)\otimes (b)\ar[r]^{(\com_{a,b})} \ar[d]_{X\otimes Y} &
  (b)\otimes (a) \ar[d]^{Y\otimes X} \\
  (c)\otimes (d) \ar[r]^{(\com_{c,d})} & (d)\otimes (c)}\\ \\
commutes for all linear transformations $X:(a)\rightarrow (c)$ and
$Y:(b)\rightarrow (d)$. Suppose
\begin{eqnarray}
(\com_{a,b})a_{i}\otimes b_{j} & = & (C_{a,b})^{kl}_{ij}b_{l}\otimes a_{k}
\end{eqnarray}
where $\{ a_{i}\}_{i}$ is a basis for $a$ and $\{ b_{j}\}_{j}$ is a
basis for $b$. Take $c=a$, $d=b$, $X=X(i;k):a_{r}\mapsto
a_{k}\delta_{i,r}$ and $Y=Y(j;l):b_{s}\mapsto b_{l}\delta_{j,s}$ in the
square diagram and apply the maps to the basis vector $a_{i}\otimes
b_{j}$. The top right half gives
\begin{eqnarray}
a_{i}\otimes b_{j} & \mapsto \sum_{m,n}(C_{a,b})^{mn}_{ij}b_{n}\otimes 
a_{m} & \mapsto (C_{a,b})^{ij}_{ij}b_{l}\otimes a_{k}
\end{eqnarray}
And the bottom left half gives
\begin{eqnarray}
a_{i}\otimes b_{j} & \mapsto  a_{k}\otimes b_{l} & \mapsto
\sum_{r,s}(C_{a,b})^{rs}_{kl}b_{s}\otimes a_{r}
\end{eqnarray}
These two being equal allows us to conclude that
$(C_{a,b})^{rs}_{kl}=\delta^{r}_{k}\delta^{s}_{l}(C_{a,b})^{ij}_{ij}$
and hence $\com_{a,b}$ can only introduce a global phase factor
$(C_{a,b})^{11}_{11}$. Moreover, if $a\cong c$ and $b\cong d$ then
$\com_{a,b}$ and $\com_{c,d}$ introduce the same phase which we denote
by $\gamma_{[a],[b]}$. That is to say the commutativity phase is
$\bA$--graded. Similar arguments allow us to conclude that all
recouplings contribute only phase factors.

\section*{Acknowledgements}                                          %

This research was supported by the New Zealand Foundation for Research,
Science and Technology. Contract number UOCX0102

\section*{References}                                                %

\end{document}